\documentstyle[12pt]{article} 
\begin{document}
\begin{center}
{\Large\bf
COMPLEMENTARITY AND PHASE DISTRIBUTIONS FOR ANGULAR MOMENTUM SYSTEMS}\\
\vskip1.50cm
{G.S. AGARWAL$^{a,}$\footnote{E-mail:~gsa@prl.ernet.in; also Honorary 
Professor at Jawaharlal Nehru Centre for Advanced Scientific Research, 
Bangalore, India.} and R.P. SINGH$^{b}$}\\ \vskip0.5cm
{$^{a}$\em Physical Research Laboratory, Navrangpura, Ahmedabad
380 009, INDIA}\\
{$^{b}$\em University of Hyderabad, Hyderabad 500 046, INDIA}\\
\vskip1.0cm
{\large\bf ABSTRACT}\\
\end{center} 
Interferences in the distributions of complementary variables for angular 
momentum - two level systems are discussed. A quantum phase distribution is 
introduced 
for angular momentum. Explicit results for the phase distributions and 
the number distributions for atomic coherent states, squeezed states and 
superpositions of coherent states are given. These results clearly 
demonstrate the issue of complementarity and provide us with results 
analogous to those for the radiation field.\\ 
\pagebreak

Complementarity of basic variables in Physics is leading to new types of 
interference experiments with wide implications [1-5]. For example 
complementarity of position and momentum variables has been used by Rauch 
and coworkers [1] to perform new types of neutron interferometric 
experiments. Wolf and coworkers [2-4] demonstrated the utility of the 
complementarity of frequency and time variables. An extension of these 
ideas to other complementary variables would be important. In particular 
one should examine systems with spins or angular momentum $\em j$. The 
excitation in the system is determined by the number distribution 
$p(m)=\langle j,m|\rho|j,m\rangle$ 
with $|m|\leq j$. Here $|j,m\rangle$ is the simultaneous eigenstate of 
$J^2$ and $J_z$ and $\rho$ is the density matrix for the system. Clearly we 
need to introduce complementary variable. This would be phase variable. 
However, there are difficulties [6] with the introduction of the phase 
operator because of the boundedness of angular momentum operator spectrum 
 $|m|\leq j$. We will avoid these difficulties by directly introducing the 
phase 
distribution $p(\varphi)$. The interferometric aspects can be discussed in 
terms of the distributions $p(m)$ and $p(\varphi)$. The analysis that we 
present is also applicable to a system of identical two level atoms, as 
such a system is equivalent to an angular momentum system [7]. Here the 
phase $\varphi$ will refer to the phase of the dipole moment of the 
system and thus $\varphi$ is an important quantity in the context of 
atomic coherences and the interferometry based on such coherences.
 
 In this letter we therefore first address the question- what are the 
phase distributions associated with angular momentum operators. Note that 
in recent years the phase distributions for the radiation field have been 
extensively discussed [8-11]. The phase distributions have been defined in a 
variety of ways as there is no unique way to do so. However most ways of 
defining are qualitatively equivalent though the different definitions 
differ in details. Under certain conditions it was demonstrated that the 
measured phase distributions [12] were related to the phase space 
distributions like the Q- function and Wigner function.
  We introduce $p(\varphi)$ via the phase space distributions [13] for the 
angular momentum operators. Let $\rho$ represent the density matrix 
\begin{equation}
\rho~=~\sum_{m,m'}\rho_{mm'}\mid j,m\rangle\langle j,m'\mid,
\end{equation}
and let $|\theta,\varphi\rangle$ be the atomic coherent state [14]
\begin{equation}
\mid \theta, \varphi\rangle~=~\sum_{m}\left[\begin{array}{c} 
2j\\ 
j+m\end{array}\right]^{1/2}\left(sin\frac{\theta}{2}\right)^{j+m}
\left(cos\frac{\theta}{2}\right)^{j-m}\mid j,m\rangle 
e^{-i\left(j+m\right)\varphi}.
\end{equation}
For this state the mean value of the dipole moment operator 
J$_{-}$=J$_{x}$~-~{\em i}$J_{y}$ is 
$jsin \theta e^{-i\varphi}$. Thus we should look for the distribution of 
$\varphi$. This is in analogy to the radiation field in the coherent 
state. A very useful phase space distribution is the Q- function defined by 
\begin{equation}
Q\left(\theta,\varphi\right)~=~\left\langle\theta,\varphi\mid\rho\mid\theta,\varphi\right\rangle,
\end{equation}
which is normalized according to
\begin{equation}
\left(\frac{2j+1}{4\pi}\right)\int\int 
Q\left(\theta,\varphi\right)\:sin\theta\:d\theta\:d\varphi~~=~1.
\end{equation}
We next define $p(\varphi)$ via
\begin{equation}
p\left(\varphi\right)~=~\left(\frac{2j+1}{4\pi}\right)\int 
Q\left(\theta,\varphi\right)\:sin\theta\:d\theta~~;~~p\left(\varphi\right)~>0. 
\end{equation}
On using (1) to (3), we find that
\begin{equation}
p\left(\varphi\right)~=~\sum_{m,m'}\rho_{mm'}~~B\left( j-\frac{m+m'}{2}+1,
~ j+\frac{m+m'}{2}+1\right) e^{i\left(m-m'\right)\varphi}, 
\end{equation}
where B(x,y) is the Beta function. Note that the number distribution is 
given 
by $\rho_{mm}$. If $\rho_{mm'}\propto\delta_{mm'}$, then 
$p(\varphi)$ is uniform as expected.

We next consider $p(\varphi)$ and $p(m)$ for some important states of 
the angular momentum systems.

A. {\bf Coherent State}~~ $\mid\alpha,\beta\rangle$

In this case the probability $p(m)$ of finding the system in the state 
$\mid j,m\rangle$ is given by 
\begin{equation}
p\left(m\right)~=~\left[\begin{array}{c}2j\\j+m\end{array}\right]
\left(sin\frac{\alpha}{2}\right)^{2j+2m}\left(cos\frac{\alpha}{2}
\right)^{2j-2m},
\end{equation}
which is just the Binomial distribution in terms of the variable $(j+m)$
 with mean value and variance equal to
\begin{equation}
\langle j+m\rangle~=~j(1-cos\alpha),~~\langle \left( 
j+m\right)^{2}\rangle~-~\langle j+m\rangle^{2}~=~\frac{j}{2}~\sin^{2}\alpha.
\end{equation}
 The corresponding phase 
distribution is 
\begin{eqnarray}
p\left(\varphi\right)&=&\left(\frac{2j+1}{4\pi}\right)\int
\langle\theta,\varphi\mid\alpha,\beta\rangle
\langle\alpha,\beta\mid\theta,\varphi
\rangle\:sin\theta\:d\theta\nonumber\\
&=&\left(\frac{2j+1}{4\pi}\right)\sum_{m=-j}^{j}\sum_{m'=-j}^{j}
\left[\begin{array}{c}2j\\j+m\end{array}\right]
\left[\begin{array}{c}2j\\j+m'\end{array}\right]\left(sin
\frac{\alpha}{2}\right)^{2j+m+m'}\left(cos\frac{\alpha}{2}\right)^{2j-m 
-m'}e^{-i\left( m-m'\right)\beta} 
\nonumber\\
& &\times \frac{2\left(j-\frac{m+m'}{2}\right)!
\left(j+\frac{m+m'}{2}\right)!}{\left(2j+1\right)!}~
e^{i\left(m-m'\right)\varphi},
\end{eqnarray}
which is centered at $\varphi$ = $\beta$.
These distributions $p(m)$ and $p(\varphi)$ are shown in the Fig.1. For 
large 
$\em j$, $p(\varphi)$ can be approximated by a Gaussian. The width of the 
distribution $p(\varphi)$ is proportional to $1/\sqrt{j}$ with 
proportionality factor $\approx $ 3.29. The width of the distribution $p(m)$ 
is proportional to $\sqrt {j}$. The two widths are thus in agreement with 
the idea of complementarity.

B. {\bf Atomic Squeezed State}~~$\mid\zeta\rangle$

The atomic squeezed state $\mid\zeta\rangle$ was defined by
\begin{equation}
\mid\zeta\rangle~=~{\cal N}~\{tanh\left( 2\mid\zeta\mid\right)\}^{J_{z}/2}~
e^{-i\frac{\pi}{2}J_{y}}\mid j,0\rangle, 
\end{equation}
where ${\cal N}$ is a normalization constant. The state $\mid\zeta\rangle$ 
has a number of very 
interesting properties. For example the distribution $p(m)$ exhibits very 
interesting interference effects [15]. This is shown in Fig.2. In terms of 
the matrix element of the rotation operator 
$d^{j}_{mp}\left(\pi/2\right)$ is 
\begin{equation} d^{j}_{mm'}\left(\frac{\pi}{2}\right)= 
\frac{\left(\left(j+m\right)!\left(j-m\right)!\left(j+m'\right)!
\left(j-m'\right)!\right)^{1/2}}{2^{j}}\sum_{q=-j}^{j}
\frac{\left(-1\right)^{q}}{\left(j-m'-q\right)!q!
\left(q+m'-m\right)!\left(j+m-q\right)!},
\end{equation}
the number and the phase distributions are shown to be
\begin{equation}
p\left(m\right)~=~{\cal N}^{2}\left(d^{j}_{m0}\left(\frac{\pi}{2}\right)
\right)^{2}~\{tanh\left(2\mid\zeta\mid\right)\}^{m},
\end{equation}
\begin{eqnarray}
p\left(\varphi\right)&=&{\cal N}^{2}\left(\frac{2j+1}{4\pi}\right)\sum_{m=-j}^
{j}\sum_{m'=-j}^{j}\left[\begin{array}{c}2j\\j+m\end{array}\right]^{1/2}
\left[\begin{array}{c}2j\\j+m'\end{array}\right]^{1/2}d^{j}_{m0}
\left(\frac{\pi}{2}\right)d^{j}_{m'0}\left(\frac{\pi}{2}\right)\nonumber\\
& &\times\{tanh\left(2\mid\zeta\mid\right)\}^{\frac{m+m'}{2}}
\frac{2\left(j-\frac{m+m'}{2}\right)!
\left(j+\frac{m+m'}{2}\right)!}{\left(2j+1\right)!}~
e^{-i\left(m-m'\right)\varphi}.
\end{eqnarray}
The unnormalized phase distribution $p(\varphi)$ is shown in the Fig.2. 
The phase 
distribution has a doublet structure which arises from the fact that $p(m)$ 
is zero for odd values of m. The peaks are at $\pm\pi/2$. This is 
because of the 
rotation by $\pi/2$ in the definition (10). This bifurcation in 
phase distribution is 
similar to the one for squeezed  states of the radiation field [10]. The 
width of each peak is proportional to $1/\sqrt{j}$ for large $\em j$ with 
proportionality factor $\approx $ 2.12. The numerical factor is less than 
that for the coherent state.

C. {\bf Superposition of Atomic Coherent States}

Finally we consider a state which is a superpositon of two atomic 
coherent states. Extensive literature exists [16] on 
superpositions of coherent states of the radiation field. For 
illustration we consider superposition of two coherent states i.e.,
\begin{equation}
\mid\psi\rangle~={\cal N}\left(\mid\pi/4,\pi/4\rangle~+~\mid\pi/4,
\pi/4+\pi/8\rangle\right), 
\end{equation}
where ${\cal N}$ is the nomalization factor. It has been shown by Agarwal 
and Puri [17] that such superpositions can be produced by considering the 
interaction of a set of atoms with a field in a cavity with large 
detuning. This dispersive interaction which is proportional to 
J$_{+}$J$_{-}$, is 
like the nonlinear phase shift term. The number and the phase distributions 
for this state are found to be 
\begin{equation}
p\left(m\right)~=~2{\cal N}^{2}\left[\begin{array}{c}2j\\j+m\end{array}\right]
\left(sin\frac{\pi}{8}\right)^{2j+2m}\left(cos\frac{\pi}{8}\right)^{2j-2m}
\left(1+cos\left(j+m\right)\frac{\pi}{8}\right),
\end{equation}
\begin{eqnarray}
p\left(\varphi\right)&=&{\cal N}^{2}\left(\frac{2j+1}{4\pi}\right)\sum_{m=-j}^
{j}\sum_{m'=-j}^{j}\left[\begin{array}{c}2j\\j+m\end{array}\right]
\left[\begin{array}{c}2j\\j+m'\end{array}\right]\left(sin
\frac{\pi}{8}\right)^{2j+m+m'}\left(cos\frac{\pi}{8}\right)^{2j-m -m'}
\nonumber\\ 
& &\times e^{-i\left(m-m'\right)\pi/4}\left(1+e^{-i\left(m-m'\right)\pi/8}
+e^{-i\left(j+m\right)\pi/8}+e^{i\left(j+m'\right)\pi/8}\right)
\nonumber\\ 
& &\times \frac{2\left(j-\frac{m+m'}{2}\right)!
\left(j+\frac{m+m'}{2}\right)!}{\left(2j+1\right)!}~
e^{i\left(m-m'\right)\varphi}.
\end{eqnarray}
These distributions (unnormalized) are shown in Fig.3. For large values of 
$\em j$, the number 
distribution $p(m)$ shows the interference minimum as given by (15) i.e.,
at {\em (j+m)}$\pi$/8~=~$\pi$ and the two peaks in $p(\varphi)$ 
become visible. The situation is similar [16] to CAT states for harmonic 
oscillator.

Thus in conclusion we have introduced a phase distribution for 
angular momentum systems. This enables us to discuss interferences in 
complementary spaces for angular momentum systems and for two level systems.

One of us (RPS) thanks National Laser Program, Government of India for 
financial support.
\pagebreak
\begin{center}
{\large \bf FIGURE CAPTIONS}
\end{center}
\begin{enumerate}
\item Phase distribution $p(\varphi)$ and number distribution $p(m)$ for
 a system in the atomic cohrent state $|\pi/4,\pi/4\rangle$ for $\em j$=10 
(dotted), 20 (dashed), and 30 (solid), the phase angle is in units of 
$\pi$.\\
\item Phase distribution $p(\varphi)$ and number distribution $p(m)$ for a 
system in atomic squeezed state $|\zeta\rangle$, where 
$\zeta$ (squeezing parameter) is equal to 2.6892, for $\em j$=10 (dotted), 20
 (dashed). For illustration $p(\varphi)$ for $\em j$=2 (long dash) is 
also drawn.\\
\item Phase distribution $p(\varphi)$ and number distribution $p(m)$ for  
superposition of two coherent states i.e., $|\pi/4,\pi/4\rangle+|\pi/4,
\pi/4+\pi/8\rangle$ for $\em j$=10 (dotted), 20 (dashed), and 30 (solid).\\
\end{enumerate}
\pagebreak
\begin{center}
{\large \bf REFERENCES}
\end{center}

[1] H. Rauch, Phys. Lett. A 173 (1993) 240; D.L. Jacobson, S.A. Werner,\\ 
\indent and H. Rauch, Phys. Rev. A 49 (1994) 3196.\\

[2] D.F.V. James, and E. Wolf, Opt. Commun. 81 (1991) 150; Phys. Lett. A 157 
(1991) 6.\\

[3] G.S. Agarwal, and D.F.V. James, J. Mod. Opt. 40 (1993) 1431.\\

[4] D.F.V. James, H.C. Kandpal, and E. Wolf, Astrophys. J. 445 (1995) 406.\\

[5] G.S. Agarwal, Found. Phys. 25 (1995) 219.\\

[6] A. Vourdas, Phys. Rev. A 43 (1991) 1564.\\

[7] G.S. Agarwal, Quantum Optics, Springer Tracts in Modern Physics, vol. 
70 (Springer, 
\indent Berlin, 1974).\\

[8] D.T. Pegg, and S.M. Barnett, J. Mod. Opt. 36 (1989) 7; Phys. Rev. A 39 
(1989) 1665.\\

[9] J.W. Noh, A. Fougeres, and L. Mandel, Phys. Rev. Lett. 67 (1991) 1426.\\

[10] W. Schleich, R.J. Horowicz, and S. Varro, Phys. Rev. A 40 (1989) 
7405.\\

[11] G.S. Agarwal, S. Chaturvedi, K. Tara, and V. Srinivasan, Phys. Rev. 
A 45(1992) 4904.\\

[12] M. Freyberger, K. Vogel, and W. Schleich, Quantum Opt. 5 (1993) 65; 
M.  Freyberger, 
\indent and W. Schleich, Phys. Rev. A 47 (1993) 30.\\

[13] The general theory of phase space distributions is described in G.S. 
Agarwal, Phys. 
\indent Rev. A 24 (1981) 2889; the illustrations of this theory 
are given in J.P. Dowling, G.S. 
\indent Agarwal, 
and W.P. Schleich, Phys. Rev. A 
49 (1994) 4101.\\

[14] F.T. Arecchi, E. Courtens, R. Gilmore, and H. Thomas, Phys. Rev. A 6 
(1972) 2211.\\

[15] For a physical explanation of this interference effect see G.S.
Agarwal, J. Mod. Opt.\\
\indent (to be published).\\

[16] W. Schleich, M. Pernigo, and F.L. Kien, Phys. Rev. A 44 (1991) 2172; V. 
Buzek,\\ 
\indent A. Vidiella-Barranco, and P.L. Knight, Phys. Rev. A 45 (1992) 6570; 
C.C. Gerry,\\ 
\indent Opt. Commun. 91 (1992) 247; K. Tara, G.S. Agarwal, and S. 
Chaturvedi, Phys. Rev. A \\
\indent 47 (1993) 5024.\\
 
[17] G.S. Agarwal and R.R. Puri, to be published.\\
\end{document}